# Radiation efficiency of heavily doped bulk n-InP semiconductor

Oleg Semyonov, Arsen Subashiev, Zhichao Chen, and Serge Luryi

*Department of Electrical and Computer Engineering,*
*State University of New York at Stony Brook, Stony Brook, NY, 11794-2350*

Recombination of minority carriers in heavily doped *n*-InP wafers has been investigated using spectral and time-resolved photoluminescence at different temperatures. Studies of the transmitted luminescence were enabled by the partial transparency of the samples due to the Moss-Burstein effect. Temporal evolution of the transmitted luminescence shows virtually no effect of surface recombination but is strongly influenced by photon recycling. Temperature dependence of the decay time suggests Auger recombination as the dominant non-radiative process at room temperature. Radiative quantum efficiency has been evaluated at different doping levels and at $2\times10^{18}$ cm$^{-3}$ it is found to be as high as 97%, which makes *n*-InP suitable for scintillator application.

## 1. Introduction

Owing to its remarkable photoluminescent and optical properties, InP is used in light emitting diodes, lasers, light detectors, solar cells, etc., and this wide-spread use has inspired extensive studies of its recombination properties [1 – 4]. High quantum efficiency combined with a relatively short decay time of luminescence makes InP a promising scintillating material for detecting high-energy particles and gamma-photons. The key issue, however, is to make a semiconductor transparent to its own interband radiation. One suggested solution [5] is to employ a heavily doped n-type semiconductor, such as *n*-InP, to produce a high Fermi level in the conduction band and thus make the bulk semiconductor partially transparent to its own luminescence. The Moss-Burstein shift [6] of the absorption edge with doping concentration provides an enlarged transparency window for the propagation of a long-wavelength portion of the luminescence spectrum through the material. Availability of epitaxial heterostructures makes it attractive to use InP as a scintillator [5,7] with the optical signal to be registered by a lattice-matched surface photo-diode of similar refractive index – ensuring efficient registration of luminescence without a significant loss due to internal reflection [8].

The published data on photoluminescence properties of heavily doped InP vary a lot and are often controversial regarding important parameters, such as the luminescence lifetime. For example, the radiative recombination constants cited in [4] and [9] differ by a factor of 6. Another important luminescence parameter, namely, the internal quantum efficiency $\beta = (1+ \nu_{nr}/\nu_r)^{-1}$, where $\nu_r$ and $\nu_{nr}$ are, respectively, the rates of recombination via radiative and non-radiative processes [10], has not been established for InP. This parameter is of crucial importance for scintillator applications.

The radiative recombination rate is sensitive to doping concentration and temperature. The observed lifetime of minority carriers is influenced by photon recycling and hence also depends on "slow" non-radiative processes such as Auger recombination, recombination on structural defects, and recombination through the deep levels. Each recombination mechanism is characterized by a specific temperature and concentration dependence. These dependences can be employed for identification of the contributing processes to the observed luminescence decay. Here we report a study of recombination processes in heavily *n*-doped InP wafers of moderate thickness of $d$=350 μm (with the Hall carrier concentrations $n$ between $2 \cdot 10^{18}$ $cm^{-3}$ and $8 \cdot 10^{18}$ $cm^{-3}$) at temperatures ranging from 77 to 330 K.

## 2. Experimental layout and results

The InP wafers supplied by Acrotec [11] were fixed to the cold finger of a cryostat (Janis Research) and irradiated by either a CW laser diode (640 nm) or a tunable (700 – 1000 nm) OPO Opolette generating 6-nanoseconds laser pulses with a repetition rate of 20 pulses per second. We shall refer to the luminescence emerging from the irradiated wafer side as the reflection luminescence and that from the opposite side as the transmission luminescence. The captured luminescence was transported through an optical fiber to either a spectroscope or a Hamamatsu PMT having 1-ns time resolution. To avoid interference with the excitation beam, the objectives collecting luminescence were placed at 20° angles to the normal capturing a solid angle of 0.0025 sr. The angular distributions of both the reflection and transmission luminescence signals were found to be close to cos $\theta$, where $\theta$ is the angle between the normal and the optical axis of the light-capturing lens.

Nevertheless, the scattered excitation light can compromise the observed luminescence signal. When the excitation energy is far above the absorption edge and beyond the luminescence spectrum, the scattered light can be filtered out. However, when the excitation laser line is within the luminescence spectrum, some contamination of the received signal is unavoidable. Fortunately, in the PMT measurements we could use temporal discrimination, since the observed decay times exceeded the laser pulse duration. As for the luminescence spectra, the relatively narrow laser line of the scattered light could be easily subtracted. To keep the excess of carrier density induced by laser radiation below the intrinsic majority carrier concentration in the samples, we used neutral filters to attenuate the laser beam.

Transmittance $T_w$ and reflectance $R_w$ of the wafers were measured over the spectral ranges of interest using a Perkin-Elmer spectrophotometer Lambda 950, which we modified to accommodate a cryostat. The transmission factor exp(-αd), the absorption coefficient α, and the reflection coefficient R were calculated from the experimental curves $T_w(T)$ and $R_w(T)$ accounting for multiple reflections from the wafers' sides, where $T$ is the temperature in K.

The luminescence pulses showed an exponential decay in the reflection channel only at excitation wavelengths in the low-intensity tails, where the noise is high and hence the accuracy low. To obtain reliable results from such non-exponential traces, one has to rely on numerical modeling of all contributing processes [3, 4, 12], which is not an easy task. The calculations are quite sensitive to the model assumptions, the most difficult of which is the correct allowance for photon recycling, and the competing processes are difficult to sort out [13]. When the excitation wavelength approaches the red wing of an absorption edge ($\alpha < 200$ cm$^{-1}$), the reflection luminescence decay curve is closer to exponential and it becomes practically indistinguishable from exponential along the total detection interval for $h\nu_{ex} \approx 1.3$ eV at room temperature. As for the transmission luminescence, the distinct exponential decay is observed in a relatively large interval even for high-energy excitation photons. In both channels, the decay time of luminescence was found to be the same if the excitation photon energy was near 1.3 eV at room temperature, i.e. when the excitation wavelength is just on the red wing of the absorption edge of the material.

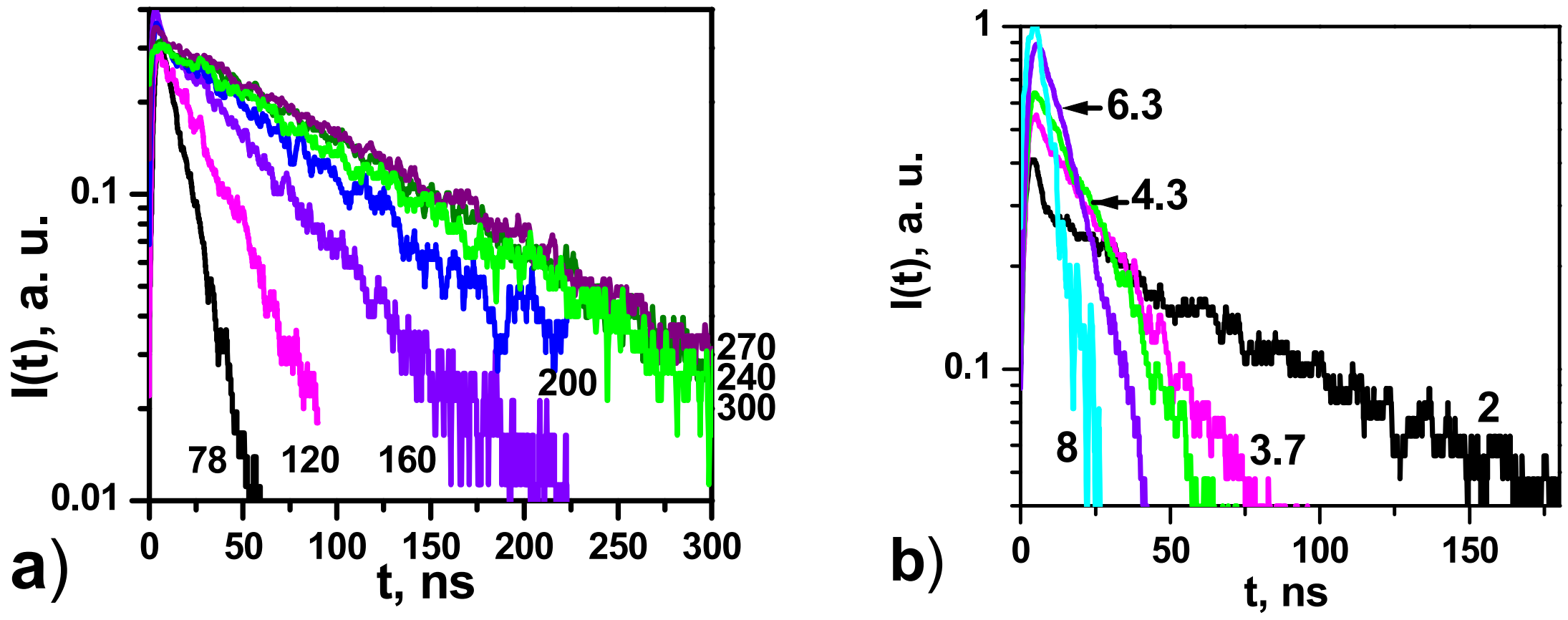

**Fig. 1** PMT signals of InP(S) transmission luminescence: (a) normalized; wafer with carrier concentration of $2{\cdot}10^{18}$ cm$^{-3}$ at different temperatures (K); (b) wafers of various doping concentration (in units of $10^{18}$ cm$^{-3}$) at a fixed excitation wavelength $h\nu_{ex} = 1.3$ eV (the time scale is expanded for better resolution of the short signals in higher concentration samples).

The luminescence intensity decay for a low-doped sample at different temperatures is shown in Fig. 1 (a). The decay becomes much faster at low temperatures. The luminescence decay time shortens with $n$ (Fig. 1b) while the amplitudes of pulses ($dI/dt$) get larger. Nevertheless, the total time-integrated luminescence emission decreases with the increasing concentration in the same way as the spectrally integrated continuous wave (CW) excitation (Fig.2a). The choice of luminescence detection method critically depends on the material properties. The lower concentration samples that produce relatively longer decay times and larger spectrally integrated signals are preferable for detection of the integrated luminescence intensity. On the other hand, the higher *n* samples produce relatively higher *dI/dt* signals and in this case time-resolved amplitude detection is preferable. In both cases, lower temperatures provide better accuracy of luminescence detection because of the higher overall luminescence intensity and higher quantum efficiency due to suppressed nonradiative recombination of minority carriers.

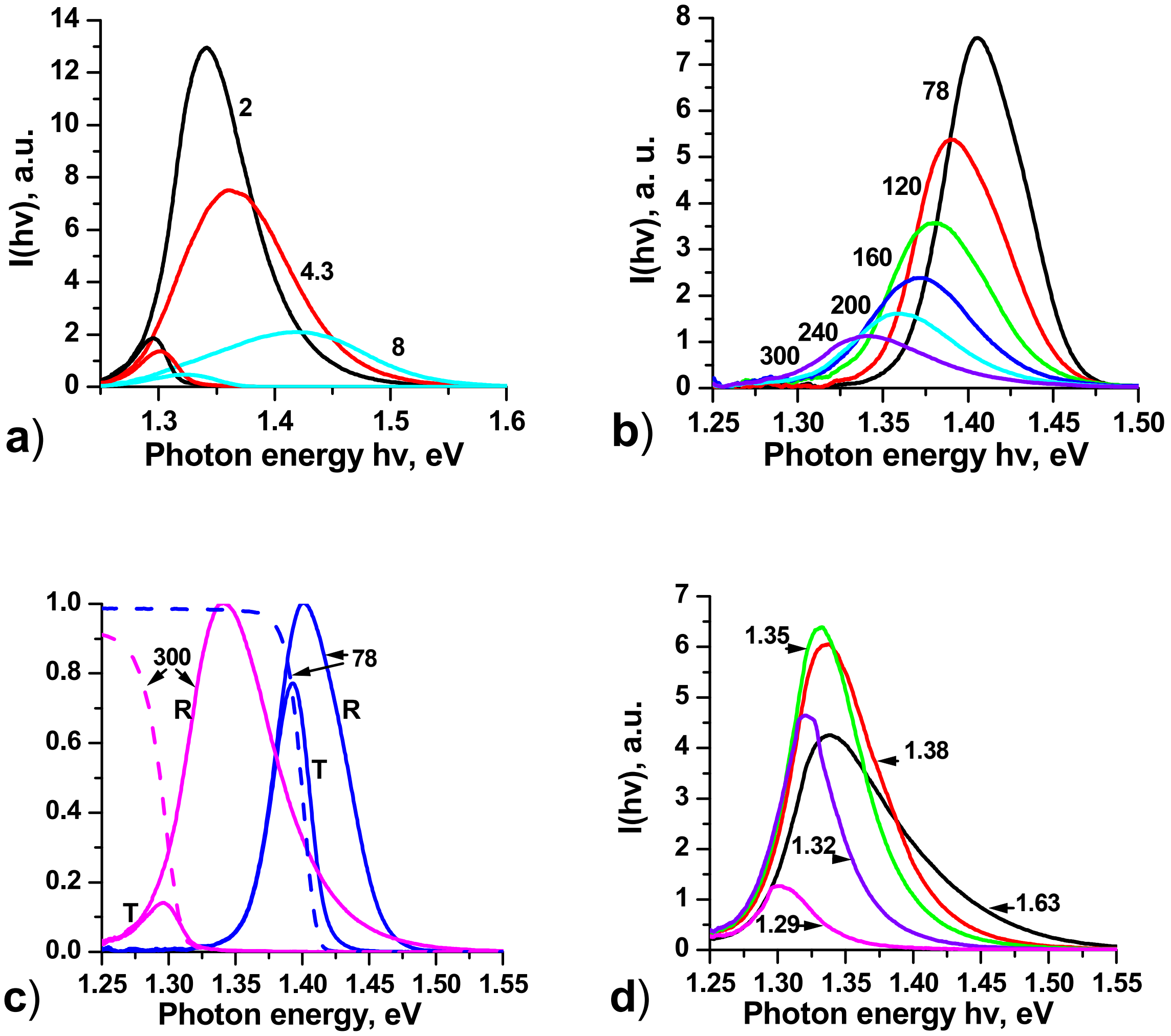

**Fig. 2** Luminescence spectra for continuous-wave (CW) excitation (a-c) at 640 nm (1.94 eV) and time-integrated OPO excitation (d). (a) Reflection (amplitude decreases with concentration) and transmission (smaller curves in the same order) spectra for different carrier concentrations *n* (in units of $10^{18}$ $cm^{-3}$); (b) evolution of the reflection spectra with temperature for $n = 2 \cdot 18$ $cm^{-3}$ (amplitude decreases with increasing temperature (K)); (c) normalized reflection (R) and transmission (T) spectra (solid curves) for $n = 2 \cdot 10^{18}$ $cm^{-3}$ at 78 K and 300 K together with the corresponding transmission factors exp(-α*d*), where $d = 353$ μm is the thickness of the wafer (dashed curves). (d) Reflection spectra from a sample with $n = 2 \cdot 10^{18}$ $cm^{-3}$ for different excitation photon energies (eV) at room temperature (OPO excitation).

The observed tendency of the reflection luminescence to become more 'transmission-like' by showing exponential decay for longer wavelength excitation correlates with the form of the emerging luminescence spectrum, as shown in Fig. 2d. When the excitation energy approaches the absorption edge, the excitation light penetrates deeper into the material and the emerging spectrum becomes more and more 'transmission-like' (compare the reflection spectrum at the excitation energy of 1.29 eV with a transmission spectrum for $n = 2{\cdot}10^{18}$ cm$^{-3}$ in Fig 2a for the excitation energy of 1.94 eV). For a large penetration depth of excitation light, the influence of surface recombination on the PMT signal vanishes while the high-energy portion of the luminescence spectrum is cut-off. Thus, measurements of the decay time become more accurate for the transmission luminescence and excitation energies at the absorption edge.

Accordingly, all the following measurements of the luminescence decay time we report from $n = 2{\cdot}10^{18}$ cm$^{-3}$ wafers were made in the transmission channel with the excitation photon energy of 1.3 eV (at room temperature; with appropriately shifted excitation photon energies at cryogenic temperatures) as shown in Fig 1a. As seen from Fig.1b, with this experimental condition all samples demonstrate exponential luminescence decay immediately after an excitation pulse and over the entire detection range. For the highest carrier concentration ($8{\cdot}10^{18}$ cm$^{-3}$) wafer (luminescence decay time $\tau \leq 6$ ns) we used for excitation a green laser emitting sub-nanosecond pulses (Teem Photonics) to provide more reliable measurements of the decay rate of transmission luminescence at lower temperatures.

As seen from the transmission luminescent spectra in Fig. 2a and c, the samples are partially transparent to the luminescence radiation. The red-side wing of the luminescence spectrum propagates through the material and emerges from the other side of the wafer. Both the reflection and transmission luminescence spectra tend to widen with the increasing carrier concentration while their relative amplitude decreases (Fig. 2a). The integrated luminescence spectra also decrease with the carrier concentration, but the ratio of spectrally integrated intensities $R = I_{tr}/I_{ref}$ increases from 6% for $n = 2{\cdot}10^{18}$ cm$^{-3}$ to 12% for $n = 8{\cdot}10^{18}$ cm$^{-3}$ at room temperature. The observed ratio dramatically increases at 78 K, where $R$ varies from 45% to 63% in the same concentration range. The relative increase of the transmission luminescence results from a blue shift of the absorption edge at low temperature (Fig. 2c) – leaving a larger part of spectrum in the low absorption region. In addition, the total intensity of luminescence increases at lower temperatures as seen from Fig. 2b.

The observed increase with $n$ of the ratio $R$ manifests the Moss-Burstein shift with the concentration. The relatively high value of $R$ suggests significant photon recycling.

Neglecting the recycling, one can calculate the transmission luminescence intensity from the experimental reflection luminescence spectrum and the experimental transmission curves corrected for multiple reflections. The intensity of thus calculated spectrum is significantly lower than that experimentally observed.

The long wavelength tails of the absorption coefficient α for different majority carrier concentrations $n$ at room temperature are shown in Fig. 3a. The relatively flat portions of the graphs just below the corresponding absorption edges show linear dependence on $n$ (Fig. 3b) and correspond, apparently, to the free-carrier Γ-L intervalley absorption [14, 15]. As noted by Casey and Stern (for $n$-GaAs samples [18]), this residual absorption below the interband absorption region appears to be sample-sensitive and depends on the chosen crystal growth technology, as well as on the degree of compensation and the concentration of residual impurities. Our experimental data (Fig. 3 b) are in agreement with [19] but differ from [14] and [15]. In particular, the residual slowly-varying absorption just below the low-energy edge of the fundamental absorption, which is believed to be caused by transitions from Γ-valley to X-valley, is nearly 1.5 times higher than that reported in [15] and 1.8 times higher than that in [14]. The observed increase of α at photon energies < 0.7 eV is typical for free-carrier absorption [15] and is close to a $\lambda^3$-dependence.

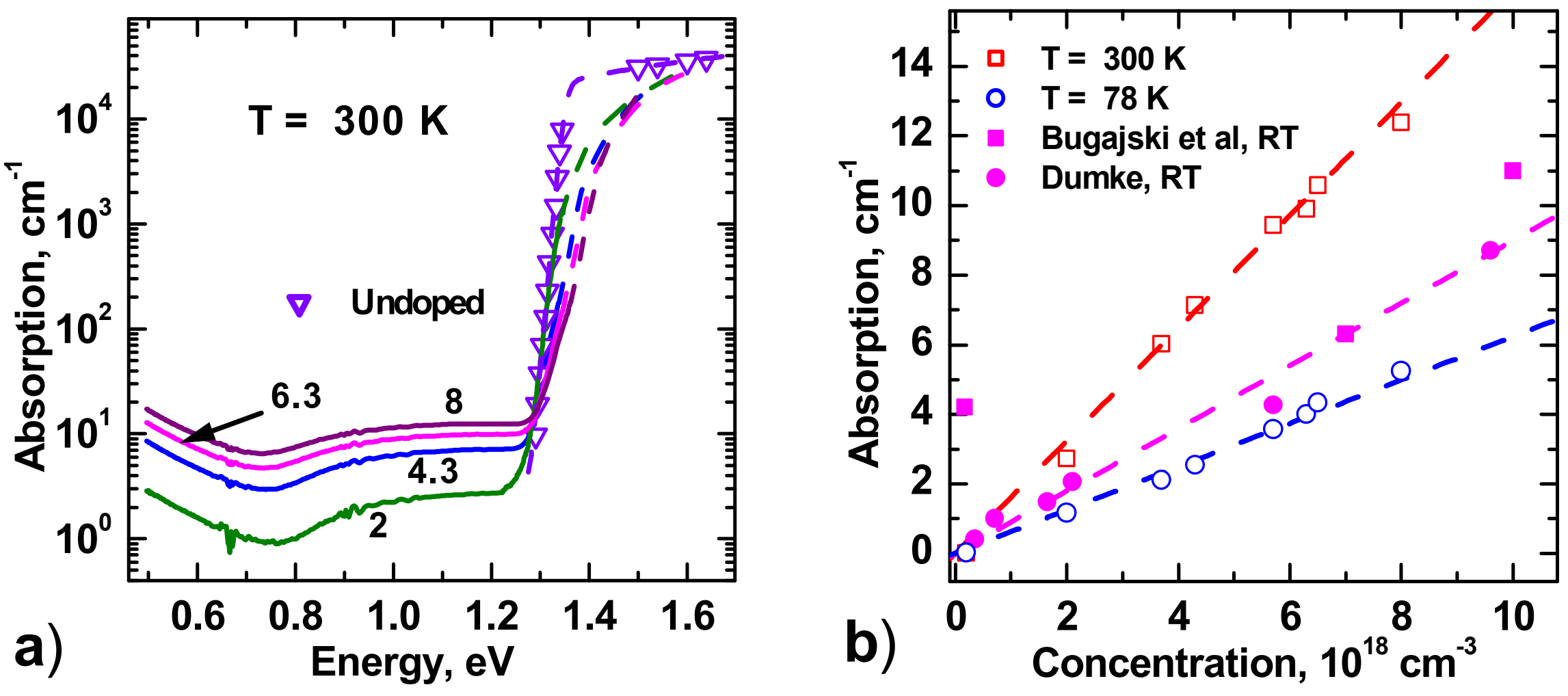

**Fig. 3** Absorption coefficient α for different photon energies and carrier concentrations $n$, $10^{18}$ cm$^{-1}$. (a) Experimental spectra (solid curves), interpolated at high energies (dash-dot curves) using the data of Ref. 14. The $2\cdot10^{18}$ cm$^{-3}$ wafer was thinned down to 50 µm enabling measurements of α up to $10^3$ cm$^{-3}$. Triangles correspond to an undoped sample (from Ref. 22). (b) Concentration dependence of the absorption coefficient at the photon energy $h\nu = 1.2$ eV.

On the other hand, the increase of the absorption coefficient at the fundamental absorption edge was found to be slightly steeper. We attribute these discrepancies to a higher compensation in our samples (within the range of 0.2-0.3), which can be estimated from the mobility data [20].

Temperature dependences of the decay rates evaluated from time-resolved luminescence detection for different concentrations are shown in Fig. 4a. The significant growth of the decay rates at lower temperatures and their increase with doping concentration is observed. The decay rate of luminescence decreases with temperature between 78 K and 200 K for all carrier concentrations studied here. It saturates near 250-300 K and then tends to grow with further increasing temperature.

## 3. Discussion

Proper interpretation of the time-resolved experimental data requires analysis of the radiative recombination rate allowing for photon recycling. Time evolution of the luminescence can be described in terms of the radiative decay rates of minority carriers with additive contributions of non-radiative mechanisms:

$$\nu = \tilde{\nu}_{\mathrm{r}} + \nu_{\mathrm{nr}} + \tilde{\nu}_{\mathrm{d}} \ , \qquad (1)$$

where $\tilde{\nu}_{\mathrm{r}}$ and $\nu_{\mathrm{nr}}$ are, respectively, the radiative and nonradiative recombination rates, and $\tilde{\nu}_{\mathrm{d}}$ is the decay rate due to minority carriers diffusion to the surface and subsequent surface recombination. The tildes denote modification of the rates by photon recycling. The radiative rate can be written as $\tilde{\nu}_{\mathrm{r}} = \nu_{\mathrm{r}} / \Phi$, where $\Phi$ is the effective recycling factor determined by the loss of luminescence photons. For very thin samples, $\Phi$ should be equal to the so-called Asbeck factor defined as an inverse probability of the photon escape through the boundaries of the luminescent region [16]. In our relatively thick samples, the photon losses are due to: (*i*) the luminescence photon escape from the excitation and observation areas; and (*ii*) the photon losses via residual (non-interband) absorption accompanied by energy dissipation [17]. For the photon losses, the dominating mechanism in *n*-type InP is the intervalley absorption by the free carriers. Therefore,

$$\frac{1}{\Phi} = \frac{1}{\varphi} + \frac{1}{\Phi_e} \ , \qquad \frac{1}{\Phi_e} = \int S(E) \frac{\alpha_e}{\alpha_i + \alpha_e} dE \ , \qquad (2)$$

where $\varphi^{-1}$ is the photon escape probability from the detection spot, $\Phi_e^{-1}$ is the photon loss probability due to residual absorption, and *S(E)* is the normalized luminescence spectrum. The absorption coefficients $\alpha_i$ and $\alpha_e$ correspond to the interband and free-carrier absorption, respectively. Both coefficients vary with the carrier concentration.

As seen from (2) the recycling factor $\Phi$ is sensitive to the form of the absorption spectra near the interband absorption edge. To evaluate the real shape of the luminescence spectrum, we used the experimental absorption curves together with their interpolation to the higher values of absorption coefficient according to [21], as shown in Fig. 3a for T=300 K. Correctness of this interpolation was tested using a wafer with $n = 2\times10^{18}$ cm$^{-3}$ thinned to $d \approx 50$ μm thus enabling measurements of the absorption coefficient up to 1000 cm$^{-1}$. For comparison, the absorption curve for undoped InP from Ref. [22] is also shown. Doped samples exhibit steeper growth of the absorption coefficient with photon energy well above the absorption edge while at the absorption edge the curves α(E) have progressively gentler sloping when carrier concentration increases.

Fig. 4 shows the temperature dependence of the luminescence decay rates measured at different concentrations. To accurately estimate the contribution of different kinetic terms to the observed decay time, one must solve the diffusion problem allowing for recycling. A comprehensive review of various analytical approaches to this problem can be found in [23]. The effect of carrier diffusion is determined by the longest of the two: the mean diffusion time to the surface and the surface recombination time [22]. Their ratio $r = s_h d / \widetilde{D}$ is determined by the surface recombination velocity $s_h$ of holes, the recycling-enhanced diffusion constant $\widetilde{D}$ and the sample thickness *d*. For the thick samples (*r* >> 1), one has $\widetilde{\nu}_d = \pi^2 d^2 / \widetilde{D}$, and in the opposite case, *r* << 1 (fast diffusion or thin sample), the decay rate is $\widetilde{\nu}_d = 2s_h / d$. Either way, the decay of the luminescence signal should be non-exponential [3, 4, 13, 23] if the minority carrier kinetics is strongly influenced by diffusion to the surface. The experimentally observed exponential decay of luminescence excited by photons with their energies near the gap energy suggests negligible contribution $\widetilde{\nu}_d$ of minority-carrier dynamics to the decay rate (Eq. 1). The high values of the recycling factor (see below) also support the conclusion that diffusion of holes to the surface is negligible; this means that the approximation *r* << 1 is valid in our case. The impact of surface recombination on the decay rate is proportional to 2*s*/d, where *s* is the surface recombination velocity, and it does not manifest itself so long as $s < 10^5$ cm/s. This upper limit of *s* seems to be in line with [3]. Besides, its contribution to the effective decay rate would be weakly increasing with concentration (via Fermi energy) and temperature (via the average carrier velocity), which contradicts to the observed variation of ν with *n* and *T* (cf. Eq. 3 below) where the linear in *n* term decreases with temperature. As for the remaining two terms in (1), their relative contribution can be estimated from the experimentally observed temperature dependences of the decay times (Fig.4, (a)) for different concentrations.

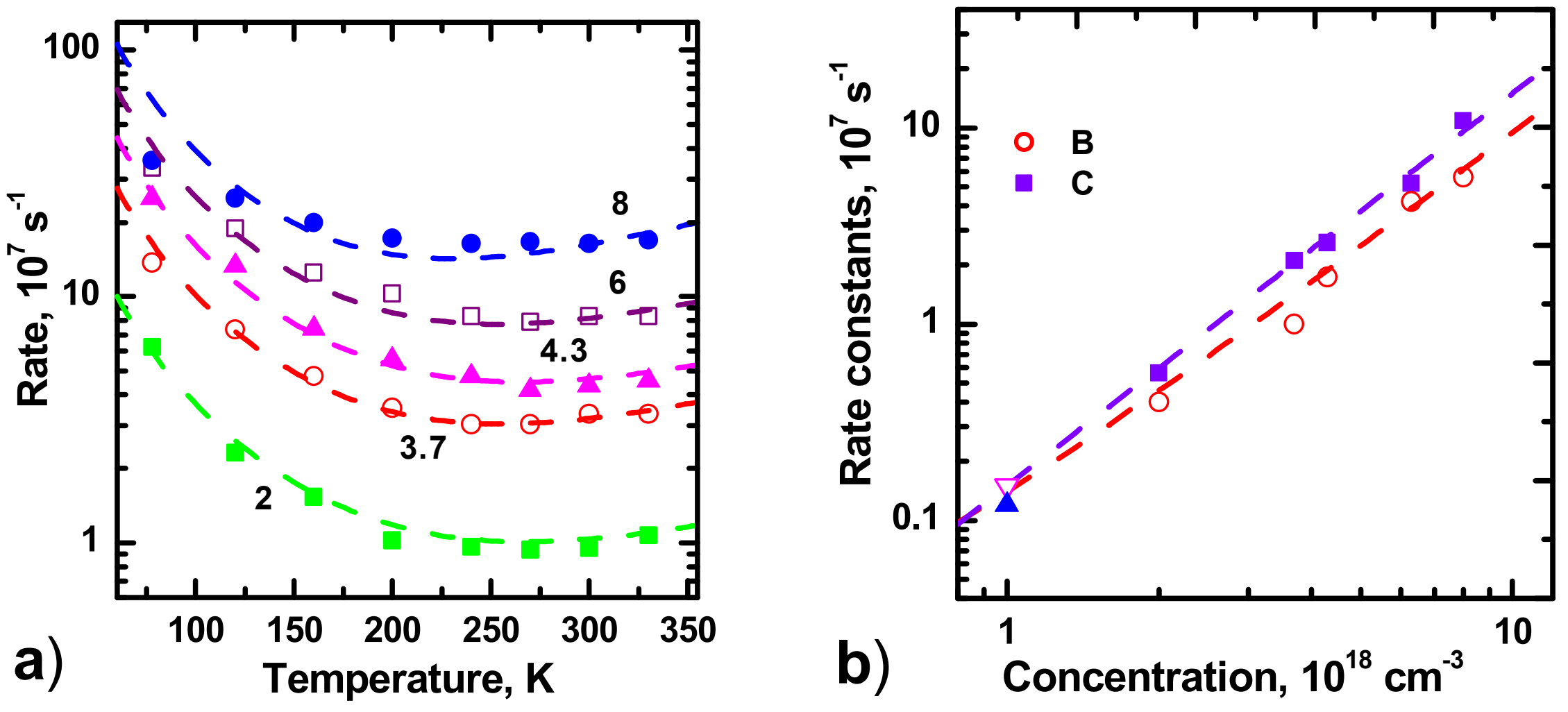

**Fig. 4** Luminescence decay rates measured by kinetic experiments and approximated by superposition of two power functions of concentration and temperature (Eq. 3). (a) Temperature dependence of the luminescence decay rate ν for different doping concentrations $n$ in units $10^{18}$ cm$^{-3}$. Dashed lines represent fits to the experimental data as in Eq. (3). For $n$ = $8{\cdot}10^{18}$ cm$^{-3}$, the experimental point at T = 78 K deviates from the approximation because of the limited detector resolution. (b) The concentration dependence of the coefficients $B$ and $C$ with approximations by quadratic polynomials. Triangles at $n$ = $1{\cdot}10^{18}$ cm$^{-3}$ correspond to the data of Ref [14].

The observed concentration dependence of the decay rate for all the samples can be approximated by the superposition of two power functions of the temperature:

$$\nu = A + \widetilde{B}(n)(300/T)^{\eta} + C(n)(T/300)^{\xi}. \qquad (3)$$

This approximation is shown in Fig. 4 (a) by the dashed lines with the best-fit parameters $A$ = $3{\cdot}10^6$ s$^{-1}$ and the power exponents η = 2 and ξ = 1. The exponents η and ξ are not determined unambiguously but the choice can be physically justified, as followed from the time-resolved data and discussion below.

The variations of parameters $\widetilde{B}(n)$ and $C(n)$ at room temperature are plotted in Fig. 4 (b). In both cases, the $n$ dependence is close to quadratic: $\widetilde{B}(n) = 0.9 \cdot 10^6 (0.55 + n/n_0)(n/n_0)$ [1/s] and $C(n) = 1.6{\cdot}10^6\,(n/n_0)^2$ [1/s], where $n_0 = 10^{18}$ cm$^{-3}$.

The first term in (3) can be plausibly interpreted as the Shockley-Read recombination rate. The 2$^{nd}$ term is the radiative recombination rate corrected for recycling. Finally, the third term can be attributed to Auger recombination because of its quadratic concentration dependence and the observed tendency to grow linearly with temperature, which is characteristic of phonon-assisted Auger recombination [24, 25] in ionic crystals. The room-temperature Auger constant $C_A = 1.5 \cdot 10^{-29}\ \mathrm{cm}^6/\mathrm{s}$ is close to cited in Chap. 12 of [25] ($1.6 \cdot 10^{-29}\ \mathrm{cm}^6/\mathrm{s}$) – estimated for the process $ehe_te$ associated with the energy transfer to electrons localized on deep donor centers. Apparently, this may also contribute to phonon-assisted Auger processes.

To determine the radiative recombination constant $B_R$, we estimated the recycling factor from Eq. (2) using the Shockley - van Roosbroek equation [6] for the emission spectrum:

$$S(E) = c\alpha_i(E)E^2 e^{-\frac{E}{kT}}, \tag{4}$$

were $c$ is the normalization constant and $E$ is photon energy. The actual interband absorption coefficient $\alpha_i(E)$ was determined subtracting the residual free-carrier absorption coefficient $\alpha_e(E)$ (displayed in Fig. 3b) from the total absorption coefficient (Fig. 3a).

Next, we estimate the photon escape probability contribution $\varphi^{-1}$ to the inverse recycling factor in Eq. (2). This contribution depends on the relative size of the excitation spot and the detection area. In our experiments, the detection area was always larger. Therefore, the escape volume $V$ can be modeled as a cylinder of height equal to the sample thickness and base equal to the detection area, so that

$$\frac{1}{\varphi} = \left\langle \frac{1}{V}\int dV\ e^{-\alpha(E)r} \right\rangle, \tag{5}$$

where $r$ is the distance from a particular luminescent point in the material to the cylinder surface and the angular brackets denote averaging over the normalized emission spectrum (4). In the evaluation of (5) we can neglect the photon escape by transmission through the flat surfaces of the cylinder, since this may happen only for the beams with their incidence angles smaller than the angle of the total internal reflection. The transmitted flux constitutes less then 2% of the total photon flux incident on one of the flat surfaces. Multiple surface reflections can be accounted for by replacing the initial escape cylinder by a cylinder of infinite height.

The resultant photon escape factor $\varphi$ together with the recycling factor $\Phi_e$ (which takes account of only free-carrier residual absorption) and the total effective recycling factor $\Phi$ are listed in Table 1 for different carrier concentrations. The effective recycling factor is plotted in Fig. 5 (squares) as function of carrier concentration at room temperature; the dashed curve shows an analytical approximation $\Phi = 78(n_0/n)$. The $n^{-1}$ dependence of the effective recycling factor suggests that *both* terms in Eq. (2) vary linearly with $n$. The room-temperature radiative recombination constant $B_R = \widetilde{B}\Phi n_0^{-1}$ is:

$$B_R = 1.2 \cdot 10^{-10} \times (1 - 0.092 \cdot n / n_0)\,\mathrm{cm}^3/\mathrm{s}\,. \quad (6)$$

At low $n$, the radiative constant $B_R$, given by Eq. (6) is fairy close to that reported in [4]. The noticeable decline of $B_R$ for $n >> n_0 = 10^{18}$ cm$^{-3}$ is similar to the $B_R$ ($n$) dependence noted for $n$-type GaAs [26].

| $n$, $10^{18}$ cm $^{-3}$ | $\varphi$ | $\Phi_e$ | $\Phi$ |
|---|---|---|---|
| 2 | 56 | 142 | 48 |
| 3.7 | 33 | 38.4 | 19.8 |
| 4.3 | 33.6 | 29 | 15.5 |
| 6 | 38 | 34 | 12.25 |
| 8 | 24 | 13 | 8.5 |

**Table 1** Photon escape factor $\varphi$ out of the detection area, free-carrier absorption factor $\Phi_e$, and the effective recycling factor $\Phi$ at T = 300 K.

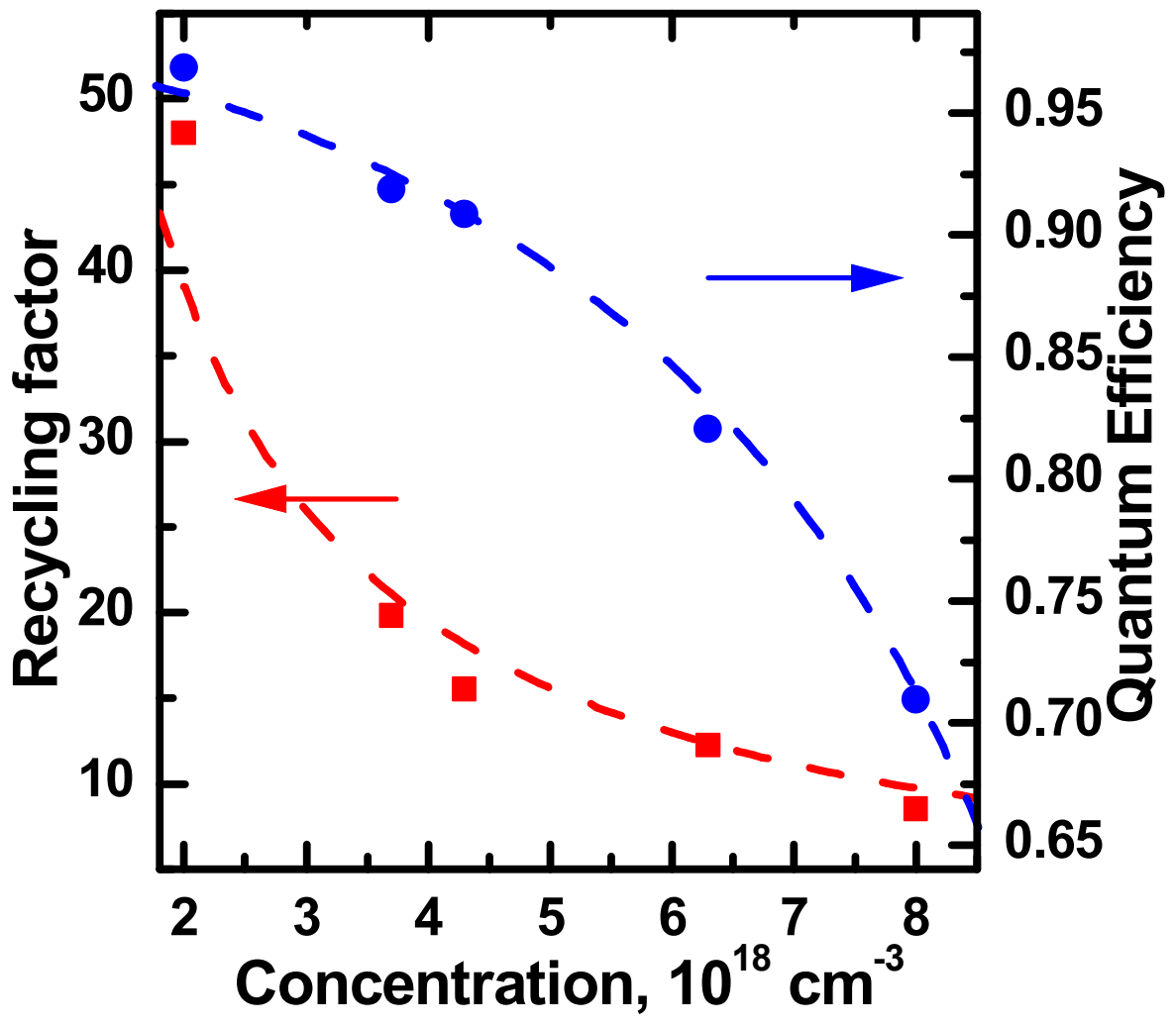

**Fig. 5** Concentration dependence of the recycling factor $\Phi$ (squares) and quantum efficiency $\beta$ (dots) obtained from the experimental data. The dashed curves are analytical approximations, $\Phi = 78(n_0/n)$ and $\beta$ in terms of coefficients $A$, $B$, $C$, and $\Phi$.

The values reported in [3, 4], where the process of photon recycling was ignored, are smaller. However, the concentration dependence of the recombination rate was not studied in [3, 4] and the temperature variation of the recombination rate was interpreted as a superposition of several recombination channels, namely the donor-to-valence-band, conduction-band-to-acceptor, and band-to-band radiative transitions. These channels manifest themselves as distinct spectral lines at liquid helium temperatures.

The radiative recombination constant $B$ is known to be proportional to $(300/T)^{\eta}$ with $\eta$ in the range between 3/2 and 2. Comparing it with the experimentally found dependence (3), it can be seen that the temperature dependence of the recycling factor appears to be weak. It would be useful to evaluate the recycling factor Φ also at 77 K to confirm the consistency of our interpretation. However, such an evaluation is not reliable, since the recombination spectrum at low temperatures is poorly described by Eq. (4). For example, equation (4) predicts an exponential slope of the spectrum edge $\sim\exp(-E/kT)$, where $E$ is the luminescence photon energy, whereas the experimentally observed slope is $\exp(-E/kT_{eff})$ with an effective temperature $T_{eff} > T$ even at low excitation power (similar observation was reported in [26]).

We can qualitatively understand the relatively slow temperature dependence of Φ as follows. One would expect that the residual absorption contribution $\Phi_e^{-1}$ in Eq. (2) should decline at low temperature. However, two factors compete here: (*i*) the observed linear decrease of the residual absorption coefficient $\alpha_e$ with temperature that causes some increase of Φ, and (*ii*) the red shift of the emission spectrum (relative to the absorption edge) which implies an increased role of residual absorption (cf. Eq. 2) and results in the increased photon losses. Apparently these two effects compensate each other.

Evaluation of the radiative recombination constant $B$ for highly doped $n$-type semiconductors is commonly based on Eq. (4) for the emission spectrum with the equilibrium rates for both absorption and emission, so that the recombination rate is estimated in terms of the absorption rate. This approach, however, is much less accurate at lower temperatures because of the exponentially strong variation of the absorption coefficient at the absorption edge. Also, the variation of the minority carrier concentration $p$ with the degenerate doping level cannot be expressed as a simple function of the intrinsic carrier concentration $n_i$ since one no longer has $n_0 p = n_i^2$. Similar arguments were used in [19] for $n$-GaAs.

Finally, we evaluate the internal quantum efficiency $\beta = (1+ \nu_{nr}/\nu_r)^{-1}$ for different concentrations (Fig. 5). It decreases from $\beta = 0.97$ for $n = 2\cdot10^{18}$ cm$^{-3}$ to $\beta = 0.71$ for $n = 8\cdot10^{18}$ cm$^{-3}$. The high values of $\beta$ for lower carrier concentrations correlate with the high ratios of spectrally integrated transmission and reflection luminescence intensities. Similar values of $\beta$ were also reported [27] for epitaxially grown GaAs layers. In contrast, significantly lower values of $\beta$ were reported for the thicker GaAs samples; it was attributed to the shorter non-radiative recombination lifetimes through the deep centers [9].

## 4. Conclusion

The moderately thick wafers of heavily doped *n*-type (S) InP have been studied using a combination of spectroscopic and time-resolved luminescence techniques in the temperature range between 78 K and 300 K. Temperature and concentration dependences of the luminescence spectra and decay rates of both the transmission and reflection luminescence signals have been obtained for various excitation wavelengths. It is shown that the most reliable measurements of the luminescence decay rates are obtained in the transmission luminescence geometry with the excitation photon energy at the low-absorption part (red wing) of the fundamental absorption edge. The observed transmission luminescence is found to be significantly enhanced by photon recycling, especially at lower doping. The photon recycling factor Φ, estimated from time-resolved experiments, increases from $\Phi \approx 8$ for $n = 8\times10^{18}$ cm$^{-3}$ to $\Phi \approx 50$ for $n = 2\times10^{18}$ cm$^{-3}$, the main limiting mechanism for low-doped samples being the photon escape from the observation spot. Similar values of the recycling factor were reported recently for *n*-doped thin GaAs layers [28]. The dominant non-radiative process is shown to be the Auger recombination. The large value of Φ results in a high internal quantum efficiency up to $\beta$ = 97% for the lowest-doped sample. This supports the important conclusion that the enhanced recycling factor and quantum efficiency are more important for the increase of $\beta$ than the Burstein shift.

Both the high quantum efficiency and the increased transparency of the relatively lower doped *n*-InP make it suitable for application in the field of radiation detection as a scintillator with high quantum efficiency and efficient photon collection. According to the experimental data, an optimal doping for scintillation applications appears to be near $10^{18}$ cm$^{-3}$ or even lower.

**Acknowledgments**

This work was supported by the Domestic Nuclear Detection Office (DNDO) of the Department of Homeland Security, by the Defense Threat Reduction Agency (DTRA) through its basic research program, and by the New York State Office of Science, Technology and Academic Research (NYSTAR) through the Center for Advanced Sensor Technology (Sensor CAT) at Stony Brook. We are grateful to Takashi Hasoda for help with thinning the samples.

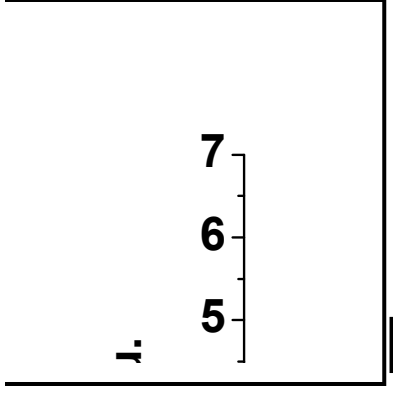